\documentclass[onecolumn]{article}
\usepackage{inputenc}
\usepackage{authblk}
\usepackage{setspace}
\usepackage[margin=1in]{geometry}
\usepackage{graphicx}
\usepackage{subcaption}
\usepackage{amsmath,amssymb}
\usepackage{lineno}
\usepackage{CJK}
\usepackage[breaklinks=true]{hyperref}
\hypersetup
{
	colorlinks=true,
	citecolor=blue,
	linkcolor=blue,
	urlcolor=blue 
}
\usepackage{natbib}
\title{\textbf{Constraining the parameters of an isolated neutron star using the lensed HI signal at uGMRT}}

\author[1*]{Rupa Basu}
\author[2]{Siddhartha Bhattacharyya}
\author[2]{Anjan Kumar Sarkar}
\author[1]{Shibaji Banerjee}
\author[3]{Debasish Majumdar}

\affil[1]{Department of Physics, St. Xavier's College, Kolkata, India, Pin-700016}
\affil[2]{National Centre for Radio Astrophysics, Pune, India, Pin-411007}
\affil[3]{Department of Physics, Ramakrishna Mission Vivekananda Educational and Research Institute, Belur Math, Howrah, India, Pin-711202}
\affil[*]{Address correspondence to: rupabasu.in@gmail.com}

\date{}

\begin{document}

\maketitle

\begin{abstract}
\noindent The strength of the HI signal originating from a distant galaxy at a cosmological distance is several orders of magnitude lower than the foreground and background noise and hence it is difficult to observe this signal at a given radio telescope.
However, a few recent studies reported the detection of that signal at the radio band suggests the strength of this signal is somehow magnified.  
In this analysis, we study the prospects of detecting this signal at different frequency bands of the uGMRT where this signal is supposed to be amplified through the strong gravitational lensing by an isolated neutron star located in a cosmological distance.
Our study shows the effects of the lensing parameters on the observables of that amplified signal and discusses its variation with the frequency bands considered here.  
We present a method to estimate the position and size of an isolated neutron star using the signal-to-noise ratio of that signal supposed to be detected at different frequency bands of the uGMRT. 
We discuss the scope of multi-messenger astronomy in the era of HI observation where the estimated lensing parameters can be cross-validated using the pulsar detection at the X-ray band from the same location in the sky. 
Our analysis is equally applicable to any radio telescope with given specifications. 
\end{abstract}


\section{Introduction}\label{sec:1}
\noindent Neutral atomic hydrogen (HI) is a fundamental component for understanding the Universe's evolution since it provides essential insights into galaxies' structure formation and dynamics (e.g. \citealt{staveley2015}; \citealt{kim2015}; \citealt{blyth2015}). 
HI signal also helps to trace the distribution of matter in the Universe.  
There are several studies indicate that redshifted HI signal in the post-reionization era ($z<6$) is more important to shed light on the structure formation (e.g. \citealt{bagla2010}) as it originates from the self-shielded dense pockets of neutral hydrogen. 
However, detecting HI signals is challenging and tactful because of its weak signal strength in comparison to the foreground/background noise, which also confines the HI observations within the local Universe. 
There are only a few observations (e.g. \citealt{chakraborty2023}; \citealt{chowdhury2020}) that extend beyond redshift $z \sim 0.1$ to provide HI signals role in more distant cosmic environments despite the limited information.

In recent decades, our grasp of the cosmic evolution of neutral hydrogen is expected to undergo a profound transformation. 
Several radio telescopes are being developed to observe the HI spectral line at high redshifts. 
These facilities include the Australian Square Kilometre Array Pathfinder \citep{johnston2008}, MeerKAT \citep{jonas2009}, the Square Kilometre Array \citep{dewdney2013}, the upgraded Karl G. Jansky Very Large Array \citep{lacy2020}, the Westerbork Synthesis Radio Telescope (WSRT) with the APERTure Tile In Focus project \citep{verheijen2008}, and the upgraded Giant Metrewave Radio Telescope \citep{gupta2017}. These instruments will perform deep HI surveys, enabling the detection of individual galaxies at redshifts of $z \sim 1$ and beyond, thus allowing for studies of HI in galaxies at unprecedented scales and distances. This increased observational capacity will require the development of new analytical frameworks that can accommodate the dynamics of an evolving Universe, ensuring precise data interpretation and advancing our comprehension of cosmic phenomena.

The HI radiation presents a significant opportunity for gravitational lensing studies. 
Several studies (e.g. \citealt{metcalf2009}) demonstrated that, if the Epoch of Reionization (EoR) occurred at redshift $z \sim 8$ or later, a large-scale radio array such as the Square Kilometre Array (SKA) could measure the lensing convergence power spectrum and refine standard cosmological parameters. 
During the EoR, the neutral hydrogen fraction is high, and HI gas is not restricted to individual galaxies. 
The HI fraction decreases significantly at lower redshifts with the gas primarily residing within discrete galaxies. 
In this lower redshift regime, the distribution of HI is modelled as discrete sources clustered according to the standard Cold Dark Matter (CDM) paradigm.  This modelling approach is essential for measuring lensing from HI radiation after reionization. The evolution of the HI fraction with redshift remains a subject of debate, significantly influencing the expected lensing signal-to-noise ratio (S/N).

In this work, we consider the HI signal originating from a distant galaxy at a cosmological distance, which gives us essential information about the properties of the neutral hydrogen source and the IGM for a given redshift $z$.
The upgraded Giant Metrewave Radio Telescope (uGMRT) (e.g. \citealt{gupta2017}) is an excellent instrument to detect that HI signal. 
The uGMRT contains a total of $30$ parabolic dishes of $45\,{\rm m}$ diameter each distributed in a "Y" shaped pattern with the longest baseline of $25\,{\rm km}$.
The uGMRT is operating in four frequency bands; Band-2 $(125-250\,{\rm MHz})$, Band-3 $(250-500\,{\rm MHz})$, Band-4 $(550-850\,{\rm MHz})$ and Band-5 $(1060-1460\,{\rm MHz})$. 
In this paper, we denote Band-2, Band-3, Band-4 and Band-5 of the uGMRT as B2, B3, B4 and B5 respectively.
We note that the information of observational frequency is reflected in the redshift of the neutral hydrogen source. 
Each frequency band has a system noise which depends on the system temperature, antenna gain, frequency bandwidth, integration time, etc. 
On top of that, the foreground radio sources of the HI signal create additional noise which is a few orders of magnitude higher than the signal.  
In general, extragalactic point sources, galactic synchrotron emission, and galactic and extra-galactic free-free emissions are considered the continuum foreground sources (e.g. \citealt{ghosh2011a}; \citealt{di2002}; \citealt{ali2008}). 
However, to tackle these foreground sources, various techniques have already been discussed in the literature (e.g. \citealt{ghosh2011b}; \citealt{datta2010}; \citealt{wang2006}; \citealt{morales2006}; \citealt{mcquinn2006}). 
The removal of these foreground sources and detection of the faint HI signal is very challenging for radio astronomers.
Recently, \citet{chakraborty2023} reported the detection of the HI signal from a neutral hydrogen source in a galaxy at redshift $z\sim1.3$ using B4.
This observation suggests that the amplitude of the signal is somehow magnified for which the signal strength is greater than the foreground noise as well as the system noise of the telescope and the HI signal was able to detect.
There are many ways to magnify the strength of the weak HI signal and the strong gravitational lensing is one of them. 
\citet{chakraborty2023} also mentioned the same and quoted the nature and the redshift of the lensing medium without giving any further details.

In the current analysis, we present a method to quantify the size and position of the lensing object using the lensed HI signal which is supposed to be detected at different frequency bands of the uGMRT. 
Many astrophysical massive objects such as MACHOs (e.g. \citealt{banerjee2003}), and black holes (\citealt{chandra}) can lens the HI signal, however here we consider the neutron star as the probable lensing object for that purpose.
Our proposed method is equally applicable to any radio telescope with a given specification. 
A brief outline of the paper is as follows: Section \ref{sec:2} presents how the lensed HI signal can be used to estimate the size and position of the lensing object. Section \ref{sec:3} discusses the results of this analysis and finally, we summarize and conclude our findings in Section \ref{sec:4}.


\section{Formalism}\label{sec:2}
In this section, we discuss the model of the strong gravitational lensing by which a neutron star (as the lensing object) can amplify the HI signal coming from a distant galaxy such that it can be detected at different frequency bands of the uGMRT.
This section is divided into several sub-sections where we discuss (1) the flux of the HI signal originating from a distant galaxy, (2) the telescope and background noise which suppress that signal, (3) the gravitational lensing model considered here for the amplification of that signal such that it can cross noise level and be detected at uGMRT, (4) the variation of the mass of a neutron star with its radius which gives a major impact on the amplification of the HI signal, (5) the optical depth which signifies the probability of detecting that signal, and (6) the chance coincidence which is an essential scenario for the amplification of the HI signal supposed to be detected at uGMRT.

\subsection{Flux of HI signal}
The brightness of any radio signal is generally quantified by the flux $S=L/4\pi D_{\rm L}(z)^2$, where $L$ is the luminosity and $D_{\rm L}$ is the luminosity distance of the source which connects with the co-moving distance by $D_{\rm L}=(1+z)\,D_{\rm M}(z)$.  
In this work, we consider the neutral hydrogen source located in a distant galaxy at a cosmological distance and $z$ is the redshift of that galaxy.
We assume that $3/4$ of the HI atoms are in the upper hyperfine state of a neutral hydrogen source, emitting the HI signal with a spontaneous emission rate of $A_{\rm HI}$. 
The number of HI atoms involved in this emission can be estimated using the luminosity of that neutral hydrogen source by the relation $n_{\rm HI}=4L/3h\,\nu_{\rm HI}\,A_{\rm HI}$, where $h$ is the Planck constant and $\nu_{\rm HI}=1420\,{\rm MHz}$ is the HI emission frequency.  
Now we introduce the HI column density which is the number of HI atoms per unit area along the line-of-sight of the distant galaxy and can be estimated using the relation $N_{\rm HI}=n_{\rm HI}/\Omega_{\rm beam}\,D_{\rm A}(z)^2$, where $\Omega_{\rm beam}$ is the solid angle covered by the telescope beam which differs for the different frequency bands of the observation and $D_{\rm A}(z)$ is the angular diameter distance of the distant galaxy (neutral hydrogen source) and it connects with the co-moving distance through the relation $D_{\rm A}(z)=D_{\rm M}(z)/(1+z)$. 
Combining all the relations discussed above, the HI flux from a distant galaxy located at a redshift $z$ can be estimated using the relation
\begin{equation}
S_{\rm HI} (z) = \frac{3\,h\,\nu_{\rm HI}\,A_{\rm HI}\,N_{\rm HI}\,\Omega_{\rm beam}}{16\pi\,(1+z)^4}
\label{eq:HI_flux}    
\end{equation}
where, $A_{\rm HI}=2.88\times 10^{-15}\,{\rm s^{-1}}$, $h=6.63\times 10^{-34}\,{\rm J\,Hz^{-1}}$ and $\nu_{\rm HI}=1420 \,{\rm MHz}$. 
\citealt{wolfe2005} argued that the damped $Ly\alpha$ system is a reservoir of the neutral hydrogen which creates the galaxies at the cosmological distance. 
The damped $Ly\alpha$ systems give the maximum contributions to the HI column density, $N_{\rm HI}$, whereas the $Ly\alpha$ systems are optically thin at Lyman limit \citep{rauch1998}. 
We consider the damped $Ly\alpha$ system as the source of HI column density and adopt the saturated value of $N_{\rm HI}=2\times 10^{20}\,{\rm cm^{-2}}$ (e.g. \citealt{wolfe2005}, \citealt{peroux2003}) for this study.
The other important parameter of the HI flux from a distant galaxy (eq.~\ref{eq:HI_flux}) is the solid angle ($\Omega_{\rm beam}$) covered by the telescope beam where $d\Omega_{\rm beam}=A(\theta)\sin\theta\,d\theta\,d\phi$ and this depends on the antenna beam pattern $A(\theta)$.
For the uGMRT, the antenna beam pattern is approximated as the Gaussian function $A(\theta)=\exp(-\theta^2/\theta_0^2)$, where $\theta_0=0.6\,\theta_{FWHM}$ and $\theta_{FWHM}$ is the angular extent of the telescope beam pattern for which the power drops by $50\%$ of its maximum value. 
We note that $\theta_{FWHM}$ varies with the different frequency bands of the uGMRT.

The redshift of the neutral hydrogen source at a distant galaxy is governed by the observational frequency $\nu_{\rm obs}$ of the telescope using the relation $z=\nu_{\rm HI}/\nu_{\rm obs}-1$.
Each frequency band of the uGMRT corresponds to a redshift range of the neutral hydrogen source, \textit{i.e.} $4.68\leq z \leq 10.83$ for B2, $1.84\leq z \leq 4.46$ for B3, $0.67\leq z \leq 1.58$ for B4 and $0\leq z \leq 0.34$ for B5 respectively.
However, in the present analysis, we consider the scenario where the HI signal originates from the self-shielded dense pockets of neutral hydrogen from a distant galaxy, which can be treated as the point source of neutral hydrogen and it is believed that this scenario happened in the post-reionization era of the evolving universe for which the redshift of the neutral hydrogen source should be $z<6$, and depending on this fact we do not consider the B2 for this work.
Further, we are interested in the detection of HI flux originating from a distant galaxy at a relatively large cosmological distance in the post-reionization era and depending on this scenario we further exclude B5 from the present analysis. 
We define a redshift called the characteristic redshift $z_c$ which is governed by the central frequency ($\nu_0$) of B3 and B4 to keep the analysis straightforward where $z_c=2.73\,{\rm and}\,1.03$ for B3 and B4 respectively.
Considering all the parameters of eq.~\ref{eq:HI_flux} mentioned above, the value of HI flux originating from a distant galaxy located at $z_c=2.73\,{\rm and}\,1.03$ are $S_{\rm HI}(z_c)=0.41$ Jy and $1.09$ Jy respectively. 
We note that the above HI flux estimation is based on \citealt{meyer2017} and the readers are referred to this paper for further details.

\subsection{Telescope and Background Noise}
The faint HI signal is attenuated by the huge telescope noise which is mainly dominated by the foreground and background noise of extragalactic point sources, galactic synchrotron emission, galactic and extra-galactic free-free emissions, etc (e.g. \citealt{ghosh2011b}, \citealt{di2002}, \citealt{ali2008}). 
The noise of any radio telescope with given parameters can be estimated using the radiometer equation \citep{lorimer2005handbook}
\begin{equation}
\Delta S = \frac{T_{sys}+T_{BG}}{N_A \,G\, \sqrt{(n_P\,T_{samp}\,\Delta \nu)}}    
\label{eq:radiometer}
\end{equation}
where, $n_P$, $N_A$ and $T_{samp}$ are the number of polarisation, number of antennae and sampling time for an observation, which are frequency band independent parameters, i.e. the value of $N_A=14$, $n_p=2$ and $T_{samp}=0.671$ sec are same for B3 and B4 respectively. 
The antenna gain $G$ varies across the frequency range of each frequency band of the uGMRT. 
However, in this analysis, we consider the minimum value of $G$ in each frequency band to maximise the telescope noise, where $G=0.38$ K Jy$^{-1}$ and $0.35$ K Jy$^{-1}$ for B3 and B4 respectively \citep{gupta2017}. 
The system temperature $T_{sys}$ is also changing across the frequency range of each uGMRT band. 
However, we consider the maximum value of $T_{sys}$ in each frequency band to maximise the telescope noise, where we consider $T_{sys}=165$ K and $100$ K for B3 and B4 respectively \citep{gupta2017}. 
The frequency bandwidth $\Delta \nu$, which is often called the usable bandwidth, is also different for the different frequency bands of the uGMRT, where $\Delta \nu=120$ MHz and $200$ MHz for B3 and B4 respectively \citep{gupta2017}.

The noise temperature ($T_{BG}$) due to foreground and background astrophysical sources also makes a significant contribution to the telescope noise estimated using eq.~\ref{eq:radiometer}.
The value of $T_{BG}$ varies with $\nu_{\rm obs}$, where it is believed that $T_{BG}$ decreases with increasing $\nu_{\rm obs}$ of the telescope. 
\citealt{bowman2018} mentioned that the value of sky temperature for the foreground and background noise of the HI signal is $\sim 6000$ K at the observational frequency of $78\pm1$ MHz using EDGES telescope, which will be even smaller if we consider B3 and B4 as the observing band to detect that signal. 
However, in this analysis, we consider a typical large value of $T_{BG}=10^4$ K same for all B3 and B4 maximise the telescope noise. 
Considering the value of noise parameters (eq.~\ref{eq:radiometer}) discussed above, the value of telescope noise $\Delta S=70.26$ Jy and $58.71$ Jy for B3 and B4 respectively which are roughly $170$ and $53$ times larger than the HI flux.
For any radio astronomical observation, we generally define a quantity called the signal-to-noise ratio, $(SNR)_{\rm HI}$, which determines how much the strength of the signal is greater than the telescope noise, which is defined as $(SNR)_{\rm HI} = S_{\rm HI}/\Delta S$.
The values of $(SNR)_{\rm HI}$ for the non-amplified HI signal are $5.9\times 10^{-3}$ and $1.9\times 10^{-2}$ at B3 and B4 respectively. 
The estimated values of $(SNR)_{\rm HI}$ mentioned above suggest that it is challenging to detect the HI signal at different frequency bands of the uGMRT without considering any signal amplification. 
However, recently \citealt{chakraborty2023} reported the $5\sigma$ detection of HI signal from a star-forming distant galaxy at a redshift of $z\sim1.3$ using B4.
They also mentioned that the HI signal is boosted by the strong gravitational lens of an early-type elliptical galaxy at redshift $z\sim0.13$. 
This observation \citep{chakraborty2023} motivates us to think about whether the strong gravitational lensing of any compact object along the line-of-sight of the source can also magnify the HI signal. 
In the present analysis, we consider the HI signal to be amplified by the neutron star for which the signal has to be magnified by a factor $\sim 10^3$ to detect that signal with a signal-to-noise ratio of $10$.

\subsection{Gravitational Lensing Model}
\begin{figure*}
\centering
\includegraphics[scale=0.25]{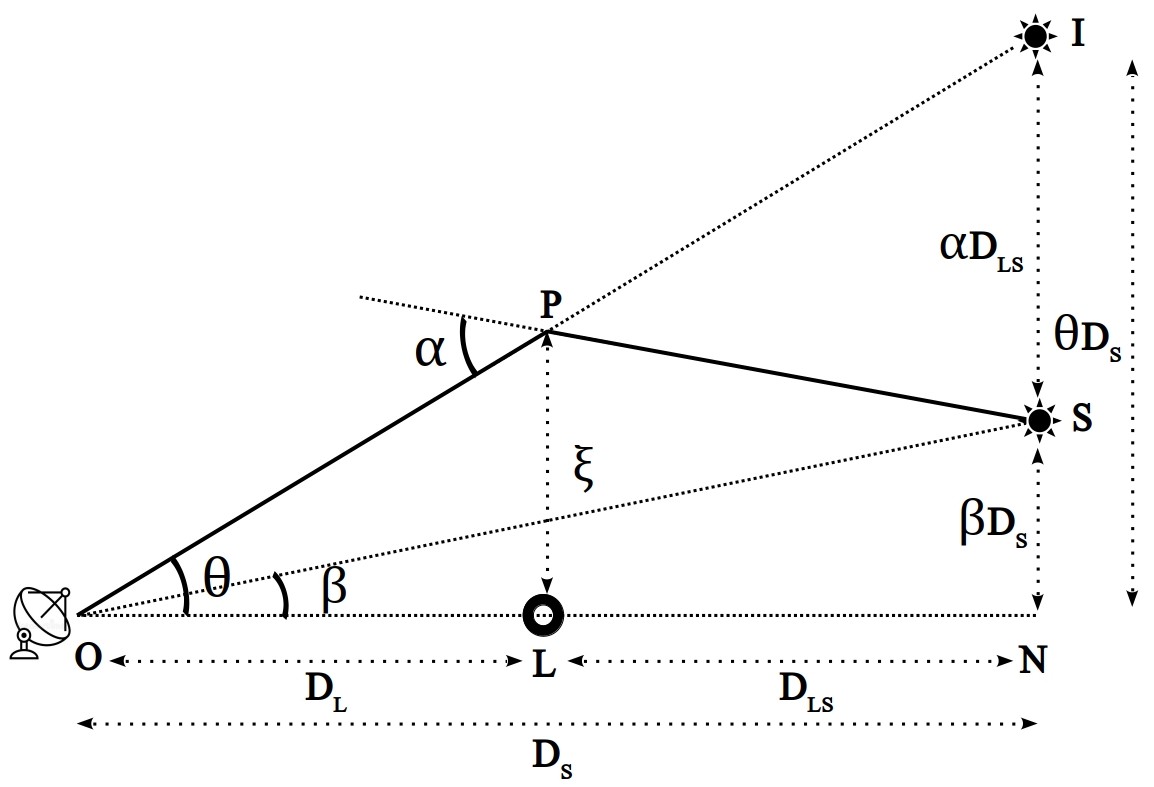}
\caption{A toy diagram for the gravitational lensing of the HI signal by an isolated neutron star located in a cosmological distance.}
\label{fig:lensing_diagram}
\end{figure*}

Figure~\ref{fig:lensing_diagram} shows a diagram that exhibits the lensing of the HI signal by a compact object which is considered to be a neutron star for this analysis. 
The source "S" emits the HI signal which initially propagates through the path "SP" and is then lensed at the point "P" due to the presence of the lensing object "L".
The observer "O" receives the signal along the path "PO" and the image of "S" is formed at "I". 
The line "ON" joining the observer "O" and compact object "L" is considered the line-of-sight (LOS) of the observation.
The angular position of source "S" and image "I" concerning "ON" are $\beta$ and $\theta$ respectively.  
It is also assumed that the image and source lie on the same plane perpendicular to "ON".
The amount of bending in the signal path is generally quantified by the parameter $\xi$ called the impact parameter which determines the closest distance of the signal path to the lensing object and $\alpha$ is the angle of bending concerning "PS".

Both source and lensing objects are located at cosmological distances and we associate the angular diameter distance with them instead of their physical distance.
For a cosmological source located at redshift $z$, the angular diameter distance $D(z) = (1+z)^{-1}\,\int_0^z dz'\,(c/H(z'))$ where $H(z)$ is the Hubble parameter and $c$ is the speed of light in vacuum. 
Referring to figure~\ref{fig:lensing_diagram}, we consider the angular diameter distance of "L" from "O" and "S" from "O" are $D_{L}$ and $D_{S}$ respectively.
The angular diameter distance of "S" from "L" is $D_{LS}$, where $D_{LS} \neq D_{S} - D_{L}$ since both "S" and "L" are located at a cosmological distance. 
The distance of the image and the source from "ON", where "ON" is perpendicular to the plane of "I" and "S", is $\beta D_S$ and $\theta D_S$ respectively whereas the distance between "I" and "S" is $\alpha D_{LS}$.  
We introduce a parameter $\gamma$, defined as $D_{L}=\gamma D_{S}$, which determines the position of the lensing object, where $\gamma \to 0$ implies the lensing object is closed to the observer and $\gamma \to 1$ implies the lensing object is closed to the source. 
We use the relation $D_{L}=\gamma D_{S}$ to determine the redshift of the lensing object $z_L$ as a function of both $\gamma$ and the redshift of the source ($z_S$), \textit{i.e.} for a fixed value of $\gamma$ and $z_S$, the value of $z_L$ is constant.  
This implies that both $D_{LS}$ and $D_L$ are the functions of $\gamma$ and $z_S$ whereas $D_S$ is the function of $z_S$ only.

Now we discuss the angular position ($\theta$) of the image "I" (figure~\ref{fig:lensing_diagram}) which is formed due to gravitational lensing of the signal from "S" (figure~\ref{fig:lensing_diagram}) by a massive compact object "L" (figure~\ref{fig:lensing_diagram}), where the value of $\theta$ can be estimated using the relation
\begin{equation}
\theta_{\pm} = \frac{\beta}{2} \left[1 \pm \sqrt{1+\left( \frac{2\theta_E}{\beta} \right)^2} \right]
\label{eq:theta}
\end{equation}
where, $\beta$ is the angular position of the source and $\theta_E$ is called the Einstein's angle. 
From eq.~\ref{eq:theta}, it is clear that two images (of the source) are formed with angular positions $\theta_+$ and $\theta_-$ concerning "ON" (figure~\ref{fig:lensing_diagram}) respectively. 
The image with angular position $\theta_+$ is formed outside of Einstein's circle centred at "S" (figure~\ref{fig:lensing_diagram}) with angular radius $\theta_E$ whereas the other image with angular position $\theta_-$ is formed inside of Einstein's circle. 
However, the observer does not detect the same flux from both images where the observer may receive more flux from the image with angular position $\theta_+$ compared to the other image. This has been discussed later in this paper. 
Einstein's angle $\theta_E$ strongly depends on the mass of the lensing object, and the angular diameter distance of the source and lensing object from the observer, which is given by
\begin{equation}
\theta_E = \left[ \frac{4 G M_L(R)}{c^2} \left(\frac{D_{LS}(\gamma,z_S)}{D_L(\gamma,z_S) \,\, D_S(z_S)}\right)\right]^{1/2}    
\label{eq:theta_E}
\end{equation}
where, $G$ is the universal gravitational constant, $c$ is the speed of light in a vacuum and $M_L(R)$ is the total mass of the lensing object depending on the radius of the object which is discussed in detail in the later part of this paper. 
The parameters $D_{LS}$, $D_L$ and $D_S$ have already been discussed earlier. 
Eq.~\ref{eq:theta_E} implies that the Einstein's angle $\theta_E$ depends on three parameters $R$, $\gamma$ and $z_S$.

We have discussed earlier that the strong gravitational lensing magnifies the flux of the HI signal due to the massive compact object "L" (figure~\ref{fig:lensing_diagram}). 
We scale the amplified flux of that signal through the relation $S_{\mu}=\mu \times S_{\rm HI}$, where $\mu$ is called the magnification factor and the value of $S_{\rm HI}$ can be estimated using the eq.~\ref{eq:HI_flux}. 
The parameter $(SNR)_{\rm HI}$ is the ratio of $S_{\rm HI}$ and $\Delta S$, where $\Delta S$ does not depend on the magnification of the HI signal by the massive compact object "L" (figure~\ref{fig:lensing_diagram}).
As a result, the parameter $(SNR)_{\rm HI}$ is also scaled similarly as we have seen for $S_{\mu}$, \textit{i.e.} the amplified signal-to-noise ratio will be $(SNR)_{\mu}=\mu \times (SNR)_{\rm HI}$.
We have discussed earlier that two images of "S" (figure~\ref{fig:lensing_diagram}) will be formed by the strong gravitation lensing due to the massive compact object "L" (figure~\ref{fig:lensing_diagram}), where one image will be formed inside and another will be formed outside of the Einstein's circle respectively.  
This implies that the magnification factor $\mu$ will also be different for the different images mentioned above. 
For a given angular position of the source "S" (figure~\ref{fig:lensing_diagram}), the value of $\mu$ can be estimated using the relation 
\begin{equation}
\mu_{\pm} = \frac{\beta^2+2\theta_E^2}{2\beta\sqrt{\beta^2+4\theta_E^2}}\pm \frac{1}{2}
\label{eq:mu}
\end{equation}
where $\mu_+$ and $\mu_-$ are the magnification factors of the images formed outside and inside of Einstein's circle respectively.
Here we can conclude that the value $(SNR)_{\mu}$ is larger for the image formed outside of Einstein's circle in comparison to another image formed there due to the strong gravitational lensing of the HI signal.   
We have seen earlier that the value of $\theta_E$ (eq.~\ref{eq:theta_E}) depends on $R$, $\gamma$ and $z_S$, and as a consequent $\mu_{\pm}$ also depends on that three parameters mentioned above. 
However, from now onwards we only consider the image of "S" (figure~\ref{fig:lensing_diagram}) formed outside of Einstein's circle for two reasons, \textit{viz.} (A) the value of $\theta_-$ is very close to $\beta$ and it may be difficult to distinguish both images simultaneously due to the small field-of-view of the uGMRT and (B) the strength of that image ($S_{\mu}$), determined through the parameter $\mu_-$, may be insufficient to detect.  
We request readers to have a quick look at the review article on gravitational lenses by \cite{narayan1996} for more details.

\subsection{Mass of a Neutron Star}
We have already discussed that Einstein's angle $\theta_E$ (eq.~\ref{eq:theta_E}) depends on the mass of the lensing object. 
As a consequence, both the parameters $\theta_+$ (eq.~\ref{eq:theta}) and $\mu_+$ (eq.~\ref{eq:mu}) also vary with the variation of the mass of the lensing object. 
Here we consider a neutron star, a compact massive object, as the lensing medium for which the mass is a non-linear function of its radius. 
The variation of mass $M_L(R)$ of a compact neutron star can be numerically evaluated by solving the Tolman-Oppenheimer-Volkov equations (TOV) \citep{oppenheimer1939}, which is given by
\begin{equation}
\frac{dP}{dR} = -\frac{G M_L(R)}{R^2}\,\rho\,\left(1+\frac{P}{\rho c^2}\right)\left(1+\frac{4\pi R^3 P}{M_L(R)c^2}\right)\left(1-\frac{2GM_L(R)}{c^2 R}\right)^{-1}
\label{eq:tov}
\end{equation}
where $P$, $\rho$ and $R$ are the pressure, density and radius of a spherically symmetric compact neutron star.  
The parameter $P$ is related with $\rho$ through a polytropic relation $P=K\rho^{\Gamma}$, where $K$ is a constant and $\Gamma$ is the polytropic exponent given by $\Gamma=1+(1/n)$ with $n$ being the polytropic index. 
The hydrostatic equilibrium inside the neutron star \citep{shapiro2008} can be described by the variation of mass $M_L$ as a function of the radial distance $R$, which is given by
\begin{equation}
\frac{dM_L}{dR} = 4\pi R^2 P
\label{eq:hydro}
\end{equation}
We estimate the value of $M_L(R)$ by solving the coupled differential equations \ref{eq:tov} and \ref{eq:hydro} numerically keeping with some suitable parameters. 
For a neutron star, which is considered here as the lensing object, we assume the value of core density $\rho_c = 2.5\times 10^{17}$ kg m$^{-3}$ and the polytropic exponent $\Gamma = 5/3$.  
The value of $M_L(R)$ increases with increasing $R$ attains a maximum value and then starts decreasing. For the neutron star with the combination of $\rho_c$ and $\Gamma$ value mentioned above, we have seen that $M_L(R)$ is maximum to the value of $\approx 2 M_{\odot}$ for $R\approx 10$ km and this is denoted here as $R_{\rm peak}$.

\subsection{Optical Depth}
Now we shift our attention to another parameter called optical depth $\tau$ which determines the detection probability of the lensed HI signal using any radio telescope with a given observational frequency. 
The parameter $\tau$ depends on the density and the position of the lensing object along the LOS of the observation and this is given by
\begin{equation}
\tau = \int_0^{z_S} d\chi(z) \, (1+z)^2 \,n_L \, \sigma(z)  
\label{eq:tau_basic}
\end{equation}
where, $\chi(z)$ is the co-moving distance of an object located at redshift $z$, $n_L$ is the co-moving number density of the lensing medium and $\sigma(z)$ is the lensing cross section of the incoming HI signal for a lensing medium of mass $M_L$ located at redshift $z$.  
Using a flat $\Lambda CDM$ cosmology and some parameter arrangements the eq. \ref{eq:tau_basic} can be written as
\begin{equation}
\tau(z_S,\gamma) = \frac{3}{2}\,\Omega_{CO}\,f \int_0^{z_S} dz \,\,\frac{(1+z)^2\,H_0^2\,D_{LS}(z,\gamma)\,D_L(z)}{c\,H(z)\,D_S(z)}
\label{eq:tau_final}
\end{equation}
where, $H(z)=H_0\,\sqrt{\Omega_M\,(1+z)^3\,+\,\Omega_{\Lambda}}$ is the Hubble parameter with $H_0$, $\Omega_M$ and $\Omega_{\Lambda}$ are Hubble constant, dark matter density parameter and dark energy density parameter respectively and the value of those parameters are taken from \citealt{planck2020}.  
The dimensionless parameter $f$ is closed to unity and the density parameter of compact objects is found to be $\Omega_{CO} \leq 0.1$ (e.g. \citealt{schneider1993}).
Here we introduced $\Omega_{CO}$ because we study the detection probability of the lensed HI signal by a compact neutron star.
From the above equation, it is clear that the value of $\tau$ solely depends on the redshift of the neutral hydrogen source ($z_S$) and the position of the lensing object which is determined by the parameter $\gamma$.

\subsection{Chance Coincidence}
Till now we have discussed how one can predict the $(SNR)_{\mu}$ of the amplified HI signal supposed to be detected at different frequency bands of the uGMRT. 
We considered the neutral hydrogen source located at redshift $z_S$ emits the signal further amplified through gravitational lensing by a compact neutron star located at redshift $z_L$. 
We have seen that the mass of the neutron star, which is a function of its radius, makes a significant contribution to this amplification and the angular position of the image, where the mass of a neutron star can be estimated using the TOV equation \citep{oppenheimer1939}. 
Apart from the mass of the lensing medium, the angular position of the neutral hydrogen source ($\beta$) also makes a significant contribution to the amplification of the HI signal, where $\beta$ is a free parameter for this analysis. 
We mentioned earlier that the angle $\beta$ is taken concerning "ON" as the LOS of the observation (figure~\ref{fig:lensing_diagram}).
From eq.~\ref{eq:mu}, we have seen that the amplification, as well as the $(SNR)_{\mu}$ of the observation, increases sharply with decreasing $\beta$, however, this is not shown in this paper. 
We have found a maximum allowed value of $\beta$, denoted as $\beta_{\rm max}$, above which we will not have sufficient amplification of the HI signal such that it will be detected at the different frequency bands of the uGMRT.
We have seen that the value of $\beta_{\rm max}$ is different for different frequency bands of the uGMRT, size and position of the lensing object. 
For a given size and position of the lensing object, i.e. $R=10$ km and $\gamma=0.5$, the value of maximum allowed angular position of the neutral hydrogen source is $\beta_{\rm max}=9.23\times 10^{-26}$ rad and $9.86\times 10^{-26}$ rad such that the HI signal can be detected at B3 and B4 respectively.
However, for different values of $R$ and $\gamma$, the value of $\beta_{\rm max}$ does not change significantly. 
All the values of $\beta_{\rm max}$ mentioned above are very small and we may conclude that $\beta_{\rm max}\rightarrow 0$ for all the frequency bands considered here. 
This implies the source has to be located close to the LOS ("ON" of figure~\ref{fig:lensing_diagram}) such that the HI signal can be detected at different frequency bands of the uGMRT. 
This is called "chance coincidence" which was already mentioned in different studies (e.g. \citealt{saini2001}) earlier.

\section{Result}\label{sec:3}
\begin{figure*}
\centering
\includegraphics[width=0.8\columnwidth]{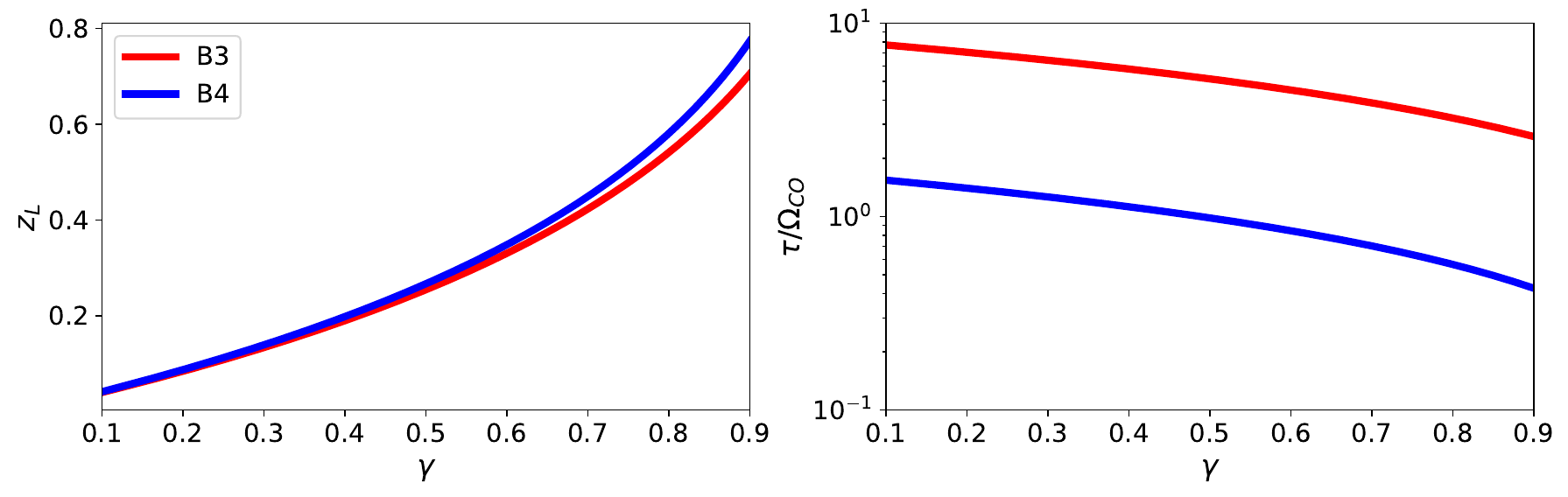}
\caption{The variation of the neutron star's redshift (left panel) and the optical depth fraction (right panel) with the neutron star's position considering the detection of the HI signal at Band-3 (red line) and Band-4 (blue line) of the uGMRT.}
\label{fig:zL_tau_gamma}
\end{figure*}

In this section, we discuss the variation of several lensing parameters and scenarios and their physical implications on the detection of the HI signal at different frequency bands (B3 and B4) of the uGMRT.
We start our discussion with the variation of the neutron star's (lensing object) redshift ($z_L$) with the parameter $\gamma$, which determines the position of the lensing object, assuming the amplified HI signal is detected at different frequency bands (B3 and B4) of the uGMRT.   
The left panel of figure~\ref{fig:zL_tau_gamma} shows the variation of $z_L$ with $\gamma$ ($0.1\leq \gamma \leq 0.9$) for the different redshifts of the neutral hydrogen source located in a distant galaxy.
Here, we use the characteristic redshift $z_c$ corresponding to different observing bands (showing in different colour lines) of the uGMRT, \textit{i.e.} $z_c=2.73\,{\rm and}\,1.03$ for B3 and B4 respectively.
We see that the value of $z_L$ increases with increasing $\gamma$ for all the frequency bands of GMRT.
For a fixed value of $\gamma$, the variation of $z_L$ with $\gamma$ for B3 (red line) and B4 (blue line) is almost similar. 
Considering all the values of $z_L$ corresponding to the complete range of $\gamma$, we find that the maximum value of $z_L$ is $0.77$ for $\gamma = 0.9$ with B4, and the minimum value of $z_L$ is around $0.04$ for $\gamma = 0.1$ which is nearly the same for both B3 and B4.
This implies that for detection of the lensed HI signal at uGMRT, the redshift of the neutron star as the lensing object lies within the range  $0.04 \leq z_L \leq 0.77$ corresponding to the different frequency bands.


Next, we discuss the variation of the optical depth with the position of the neutron star (lensing object).
The parameter optical depth ($\tau$) determines the detection probability of the amplified HI signal, where $\tau$ depends on several factors, \textit{viz.} redshift ($z_S$) of neutral hydrogen source, position ($\gamma$) of neutron star along LOS and density parameter ($\Omega_{CO}$) of compact objects. 
We do not have a clear picture of the value of $\Omega_{CO}$, however, \citealt{schneider1993} put a constraint on the upper limit of $\Omega_{CO}$ ($\leq 0.1$).
Here we study the variation of $\tau/\Omega_{CO}$ with the variation of $\gamma$ for different source redshifts where these are assumed to be the characteristic redshifts corresponding to different frequency bands of the uGMRT.  
The right panel of figure~\ref{fig:zL_tau_gamma} shows the variation of $\tau/\Omega_{CO}$ with $\gamma$ ($0.1\leq \gamma \leq 0.9$) for different redshifts of the neutral hydrogen source.
The red and blue lines show the variation of $\tau/\Omega_{CO}$ for the detection of amplified HI signal at B3 ($z_c=2.73$) and B4 ($z_c=1.03$) respectively.
For all the frequency bands considered here, we see $\tau/\Omega_{CO}$ decreases with increasing $\gamma$.
This implies that the probability of detecting amplified HI signals at different frequency bands is large if the neutron star is close to the observer.
For a particular value of $\gamma$, we find the value of $\tau/\Omega_{CO}$ is highest for B3 and lowest for B4.  
This points to the fact that we have a good amount of detection probability of the amplified HI signal if the neutral hydrogen source is far away from the observer and this is relatively poor for the scenario where the neutral hydrogen source is nearby. 
Considering all the values of $\tau/\Omega_{CO}$ corresponding to the complete range of $\gamma$, we find that the maximum value of $\tau/\Omega_{CO}$ is $7.72$ for $\gamma = 0.1$ with B3 ($z_c = 2.73$), and minimum value of $\tau/\Omega_{CO}$ is $0.43$ for $\gamma = 0.9$ with B4 ($z_c=1.03$). 
Considering the upper limit of $\Omega_{CO}$ \citep{schneider1993}, we find the maximum and minimum values of optical depth are $\tau_{\rm max}=0.77$ and $\tau_{\rm min}=0.04$. 
Combining all the information mentioned above, we may conclude that the probability of detecting amplified HI signal is maximum at B3 ($z_c = 2.73$) if the neutron star (lensing object) is close to the observer. 


\begin{figure*}
\centering
\includegraphics[width=0.8\columnwidth]{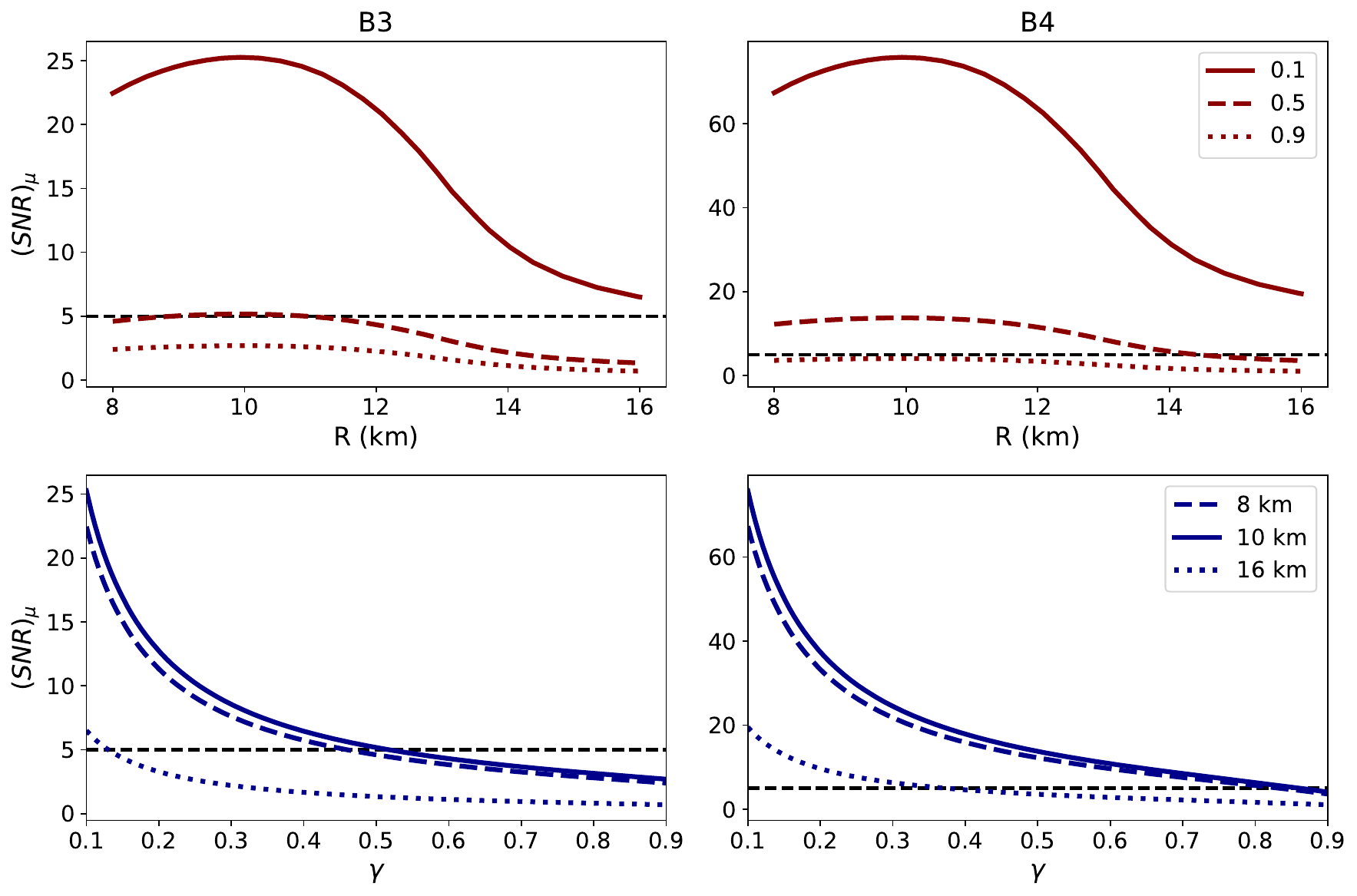}
\caption{The effect of radius (upper panels) and position (lower panels) of an isolated neutron star on the signal-to-noise ratio $(SNR)_{\mu}$ of the HI signal supposed to be detected at Band-3 (left column) and Band-4 (right column) of the uGMRT. The black horizontal dashed line shows the threshold value of $(SNR)_{\mu}$, \textit{i.e.} $SNR_{\rm Th}=5$.}
\label{fig:SNR_radius_gamma}
\end{figure*}

The amplification strength of the HI signal through the gravitational lensing by an isolated neutron star mainly depends on two physical parameters of the lensing object, \textit{viz.} its mass and position along LOS.
Previously, we have discussed that the neutral hydrogen source has to be located close to LOS, \textit{i.e.} $\beta\rightarrow 0$, such that the amplified HI signal can be detected at different frequency bands of the uGMRT which is called chance coincidence. 
For the fact stated above, we exclude the parameter $\beta$ in the phenomena of the amplification of HI signal. 
Further, we have also seen that the mass of an isolated neutron star strongly depends on its radius and that can be estimated using the TOV equation (eq.~\ref{eq:tov}).
We have converted the amplification strength into observed $(SNR)_{\mu}$ which is supposed to be detected at different frequency bands of the uGMRT.
Therefore, we may conclude that the observed $(SNR)_{\mu}$ strongly depends on the size and position of an isolated neutron star along LOS.

Figure~\ref{fig:SNR_radius_gamma} shows the variation of $(SNR)_{\mu}$ with the variation of neutron star radius and position along LOS assuming the amplified HI signal is detected at B3 (left column) and B4 (right column).
The upper panels show the variations of $(SNR)_{\mu}$ with neutron star radius ($R$) for its different positions ($\gamma$) along LOS where we consider three distinct values of $\gamma$, \textit{i.e.} $0.1$, $0.5$ and $0.9$, which spans the entire range of $\gamma$.
The lower panels show the variations of $(SNR)_{\mu}$ with neutron star positions ($\gamma$) for its different radius ($R$) values where we consider three distinct values of $R$, \textit{i.e.} $8$ km, $10$ km and $16$ km, which spans the entire range of radius.
Here we introduce a threshold value for $(SNR)_{\mu}$, \textit{i.e.} $SNR_{\rm Th} = 5$, such that any detection of the HI signal with $(SNR)_{\mu}<SNR_{\rm Th}$ will not be considered as a genuine detection which is shown in the black horizontal dashed lines in the figure~\ref{fig:SNR_radius_gamma}.

First, we discuss the variation of $(SNR)_{\mu}$ with $R$ for the detection of amplified HI signal at different frequency bands of the uGMRT.
For B3 ($z_c=2.73$) with all the values of $\gamma$ shown in the left upper panel, $(SNR)_{\mu}$ increases with increasing $R$ attains a maximum value near $R_{\rm max}\approx 10$ km and then starts decreasing with further increasing of $R$. 
However, the sharpness of the plot changes with different values of $\gamma$, where this is steeper for $\gamma=0.1$ and relatively flatter for $\gamma=0.9$. 
Further, the peak value of $(SNR)_{\mu}$ also changes with different values of $\gamma$ where this is maximum, \textit{i.e.} $SNR_{\rm max}=25.26$, for $\gamma=0.1$ and this is minimum, \textit{i.e.} $SNR_{\rm min}=2.69$, for $\gamma=0.9$.
For different values of $\gamma$ with $8\,{\rm km}\leq R\leq 16\,{\rm km}$, the values of predicted $(SNR)_{\mu}$ have not been always above $SNR_{\rm Th}$, rather for some of the values of $\gamma$, it goes below the threshold value. 
Considering $8\,{\rm km}\leq R\leq 16\,{\rm km}$, we find that $(SNR)_{\mu}>SNR_{\rm Th}$ for $\gamma=0.1$ and $(SNR)_{\mu}<SNR_{\rm Th}$ for $\gamma=0.9$, whereas for $\gamma=0.5$, $(SNR)_{\mu}>SNR_{\rm Th}$ is possible only if $R\lesssim 11$ km.
The similar qualitative natures of $(SNR)_{\mu}$ vs $R$ plots have been seen for B4, where the allowed range of $R$ for different values of $\gamma$, and the values of $SNR_{\rm max}$ and $SNR_{\rm min}$ are different.
For the detection of amplified HI signal at B4 with characteristic redshift $z_c=1.03$ shown in the right upper panel of figure~\ref{fig:SNR_radius_gamma}, we find that $(SNR)_{\mu}>SNR_{\rm Th}$ for $\gamma=0.1$ with $8\,{\rm km}\leq R\leq 16\,{\rm km}$ and for $\gamma=0.5$ with $R \lesssim 14\,{\rm km}$, whereas the predicted values of $(SNR)_{\mu}$ are always below $SNR_{\rm Th}$ for $\gamma=0.9$ with the full range of $R$ considered here. 
Here, the value of $SNR_{\rm max}$ and $SNR_{\rm min}$ are $75.74$ and $4.06$ respectively.

Now, we discuss the variation of $(SNR)_{\mu}$ with $\gamma$ for the detection of amplified HI signal at different frequency bands of the uGMRT.
For B3 ($z_c=2.73$) with all the values of $R$ shown in the left lower panel, $(SNR)_{\mu}$ decreases with increasing $\gamma$.
The qualitative nature of $(SNR)_{\mu}$ vs $\gamma$ plot is almost similar for both $R=10$ km $8$ km, however, this is largely differed for $R=16$ km.
However, for a fixed value of $\gamma$, the predicted $(SNR)_{\mu}$ is maximum for $R=10$ km and minimum for $R=16$ km and for that reason, $R\approx10$ km is denoted here as $R_{\rm peak}$, \textit{i.e.} a particular value of neutron star's radius for which the signal amplification is maximum.   
We find the predicted values of $(SNR)_{\mu}$ are not always above $SNR_{\rm Th}$ for the range of $\gamma$ considered here which further gives us the upper limit of $\gamma$, denoted by $\gamma_{\rm cut}$, and this varies with different values of $R$ mentioned in the figure. 
We see $\gamma_{\rm cut}\approx 0.46$, $0.51$ and $0.13$ for $R=8$ km, $10$ km and $16$ km respectively.
The similar qualitative natures of $(SNR)_{\mu}$ vs $\gamma$ plots have been seen for B4 of the uGMRT, where the value of $\gamma_{\rm cut}$ differs for different values of $R$  here.
For the detection of amplified HI signal at B4, we find $\gamma_{\rm cut}\approx 0.84$, $0.86$ and $0.37$ for $R=8$ km, $10$ km and $16$ km respectively.
In short for a fixed value of $R$, $\gamma_{\rm cut}$ increases with decreasing redshift of the neutral hydrogen source located in a distant galaxy.

Combining the information discussed above we may conclude that a neutron star with radius $8\,{\rm km}\leq R\leq 16\,{\rm km}$ located close to the observer can amplify the HI signal originating from a neutral hydrogen source located in a distant galaxy with redshift within the range $0.13\leq z_S\leq 2.73$ such that it can be detected at different frequency bands of the uGMRT. 
This can also happen for the other position of the neutron star along LOS with an upper limit on its radius.
Previously we have discussed the variation of a neutron star's mass with its radius governed by the TOV equation (eq.~\ref{eq:tov}) and the same feature has been replicated in the variation of $(SNR)_{\mu}$ with $R$ because the mass of a neutron star as a whole gives a major contribution to the amplification of any astrophysical signal through gravitational lensing. 
The value of $SNR_{\rm max}$ increases with the decreasing redshift of the neutral hydrogen source because the HI flux ($S_{\mu}$) from that source is large if the source is located nearby in comparison to the source which is relatively away from the observer and also the telescope noise is minimum at B4 and maximum at B3.  
Further, the value of $\gamma_{\rm cut}$ decreases with the increasing of neutral hydrogen source redshift which implies if the source is away from the observer, a neutron star with a fixed radius (say $10$ km) has to be located close to the observer to get sufficient amplification of the HI signal such that it can be detected at uGMRT.
For all of the results discussed above we have seen that a neutron star with a radius $10$ km, denoted by $R_{\rm peak}$, amplifies the HI signal in such a way that the predicted $(SNR)_{\mu}$ is always the maximum, which is irrespective of the neutral hydrogen source redshift. 
From now on, we consider $R_{\rm peak}$ as the standard radius of a neutron star for further discussion.


\begin{figure*}
\centering
\includegraphics[width=0.8\columnwidth]{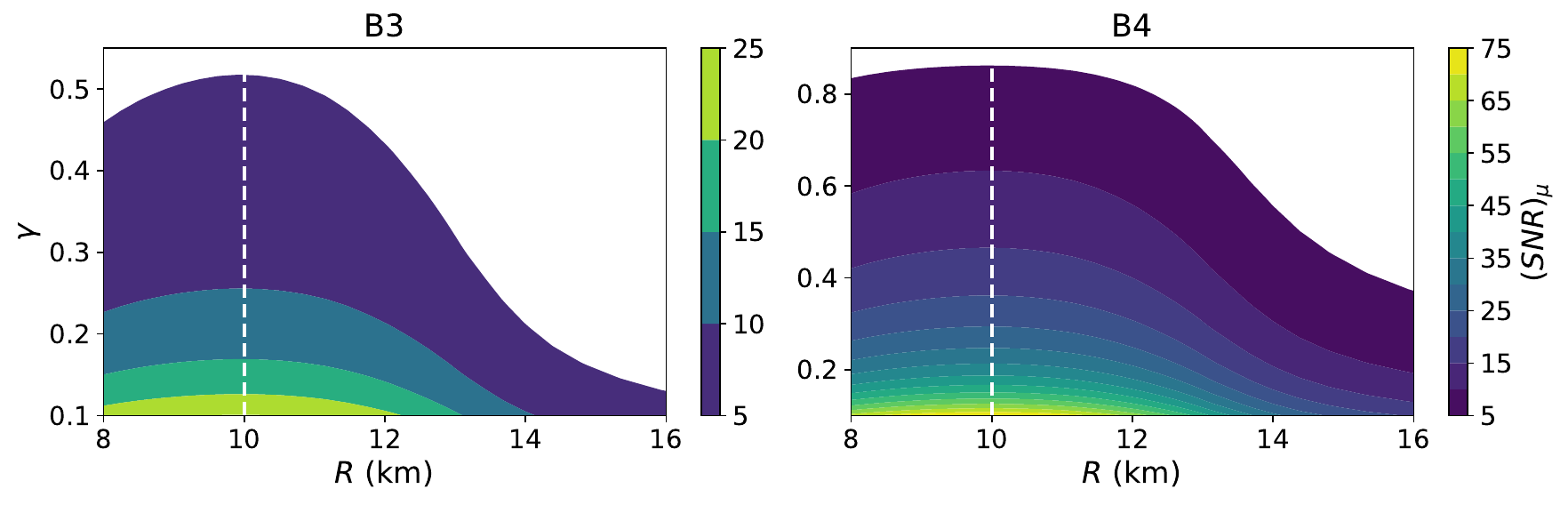}
\caption{The estimation of the radius and position of an isolated neutron star using the signal-to-noise ratio $(SNR)_{\mu}$ of the HI signal supposed to be detected at Band-3 (left column) and Band-4 (right column) of the uGMRT. The white vertical dashed line shows the standard radius of that neutron star, \textit{i.e.} $R_{\rm peak}=10\,{\rm km}$.}
\label{fig:SNR_Contours}
\end{figure*}

Till now we have analysed how the lensing parameters, such as the position and radius of a neutron star, control the amplification of the HI signal which is supposed to be detected at different frequency bands of the uGMRT. 
Now we discuss if an amplified HI signal is detected at a particular frequency band of the uGMRT with a given $(SNR)_{\mu}$ then how we can provide constraints on the radius and position of a neutron star which amplifies that signal. 
Figure~\ref{fig:SNR_Contours} shows the variation of $(SNR)_{\mu}$ on $\gamma-R$ plane for different frequency bands of the uGMRT which are supposed to be used to detect that HI signal. 
Previously we have discussed the parameter $\gamma_{\rm cut}$ which varies with $R$ and different frequency bands and the same feature is also exhibited here. 
Further, the parameter $R_{\rm peak}$, which is the standard radius of a neutron star, has been shown in the figure by a white vertical dashed line.  
Let us first consider B3 (left panel), if an amplified HI signal ($z_c=2.73$) is detected with $(SNR)_{\mu}=10$, we may conclude that a neutron star with a radius $R_{\rm peak}\approx 10$ km located at a position $\gamma=0.25$ amplifies that signal. 
However, the value of $\gamma$ may change if that neutron star has a different radius rather than $R_{\rm peak}$.
For a given $(SNR)_{\mu}$, the value of $\gamma$ increases with increasing $R$, attains a maximum value and then starts decreasing with further increase of $R$. 
Therefore it may be concluded that we have a maximum allowed value of $\gamma$ which in general corresponds to $R_{\rm peak}$ for a given $(SNR)_{\mu}$ of the signal. 
We can make the same kind of prediction about the position and radius of a neutron star if that amplified HI signal is detected at B4 (right panel). 
Suppose a HI signal is detected at B4 ($z_c=1.03$) with $(SNR)_{\mu}=10$, then that signal can be amplified by a neutron star with $\gamma=0.63$ and $R=R_{\rm peak}\approx 10$ km. 
Briefly, we may conclude that if an HI signal is detected at a particular frequency band of the uGMRT with a given $(SNR)_{\mu}$, we can predict the position and radius of an isolated neutron star which may amplify that signal. 
However, the position ($\gamma$) and radius ($R$) of that neutron star are highly correlated, but we can put a tight constraint on the value of $\gamma$ if we consider the standard radius $R=R_{\rm peak}$ for that neutron star.

\section{Discussion}\label{sec:4}
The detection of the HI signal from a distant galaxy located at a cosmological distance using a given telescope is a challenging part of radio astronomy since the strength of that signal is several orders of magnitude lower than the foreground/background noises. 
However, recently, several studies (e.g. \citealt{chakraborty2023}) reported the detection of that signal at different radio telescopes with a sufficiently large amount of signal-to-noise ratio, which implies that the strength of that signal is somehow magnified. 
There are several ways to magnify the HI signal and strong gravitational lensing is one of them. 
In this scenario, we have presented a method to estimate the position and size of the lensing medium using the signal-to-noise ratio of the amplified HI signal supposed to be detected at different frequency bands of the uGMRT considering the fact an isolated neutron star amplifies the strength of that HI signal.  

In this paper, we have studied the effects of several lensing parameters, \textit{viz.} size, mass and position of an isolated neutron star, detection probability, etc., on the signal-to-noise ratio of the HI signal and its variation with frequency bands, where Band-3 and Band-4 of the uGMRT have been considered for this analysis.
Our analysis shows there is a high possibility of detecting the HI signal from a distant galaxy using the uGMRT with a large amount of signal-to-noise ratio, which is relatively high for Band-4 in comparison to Band-3.
For the sufficient amplification of that HI signal, the neutral hydrogen source has to be located close to the line of sight that connects the observer and the isolated neutron star, where the redshift of this isolated neutron star should lie within the range $0.04 \leq z_L \leq 0.77$ corresponds to different redshifts of the neutral hydrogen source.
Further, the amplification of that HI signal is maximum for an isolated neutron star with radius $R=10\,{\rm km}$ and minimum for $R=16\,{\rm km}$ irrespective of its position along the line of sight.
We have also found an upper limit on the neutron star's position above which it can not perform sufficient amplification of the HI signal, where this upper limit varies with its radius and the redshift of the neutral hydrogen source. 
The major application of our analysis is to find out the position and size of an isolated neutron star using the signal-to-noise ratio of the amplified HI signal supposed to be detected at Band-3 and Band-4 of the uGMRT, where the neutron star's position can be estimated considering a fixed value of its radius ($R=10\,{\rm km}$), on the other way, one can also determine its radius if its position can be located by some other surveys.

Now we discuss the cross-validation of the estimated position and size of an isolated neutron which magnifies the strength of the HI signal originating from a distant galaxy, where it is believed that an isolated neutron star is one of the sources of pulsar emission.
In this scenario, we propose to cross-validate the estimated position and size of that isolated neutron star through the pulsar observation in the same location in the sky from where that amplified HI signal is supposed to be detected.
Considering the predicted redshift range of that isolated neutron star mentioned above, the required luminosity of that extragalactic pulsar emission is the order of $\sim 10^{32}\,{\rm erg\,s^{-1}}$, which is several orders of magnitude higher than the brightest pulsar (e.g. villa, Crab, etc.) detected at the radio band to date. 
This implies that it is difficult to detect the pulsar signal and the amplified HI signal simultaneously at any frequency band of the uGMRT. 
However, a pulsar signal with that required luminosity can be detected at the X-ray band (e.g. \citealt{walter1996}).
In this scenario, we discuss the scope of multi-messenger astronomy, where that amplified signal may be detected at the radio band and the corresponding pulsar signal can be detected at the X-ray band from the same location in the sky.   
Several studies (e.g. \citealt{pons2002}; \citealt{kaplan2007}) show how one can estimate the redshift and radius of an isolated neutron star which emits an extragalactic pulsar signal at the X-ray band. 
In this regard, we propose to detect the amplified HI signal at the radio band and cross-validate the estimated position and size of an isolated neutron star using the pulsar observation at the X-ray band from the same location in the sky.

We have discussed the cross-validation of the estimated lensing parameters using the pulsar observation but if that pulsar signal has not been detected along with that amplified HI signal from the same location in the sky, this implies that the HI signal may not be amplified by an isolated neutron star where some other massive compact objects can perform this amplification.  
Further, we have considered an isolated neutron star as a lensing medium, however, an accreting neutron star in a binary system can also perform the same kind of amplification which is beyond the scope of this analysis and we have a plan to study this prospect in our subsequent paper.

\bibliography{reference}

\end{document}